\RequirePackage{fixltx2e}
\documentclass[aps,pre,twocolumn,showpacs,superscriptaddress,10pt]{revtex4-1}

\usepackage[normalem]{ulem}
\usepackage{xcolor}

\usepackage{bm}
\usepackage{graphicx}
\usepackage{latexsym}
\usepackage{amsmath}
\usepackage{hyperref}
\usepackage{pxfonts}
\usepackage{cancel}
\usepackage{bbding}
\usepackage{subfig}
\usepackage{float}
\usepackage{subfloat}
\usepackage{caption}
\usepackage[section]{placeins}
\usepackage{csquotes}
\DeclareGraphicsExtensions{.pdf,.png,.jpg}

\begin{document}

\title{Optimal Positron-Beam Excited Plasma Wakefields in Hollow and Ion-Wake Channels}

\author{Aakash A. Sahai} 
\email{aakash.sahai@gmail.com}
\affiliation{Department of Electrical Engineering, Duke University, Durham, NC, 27708 USA}
\author{Thomas C. Katsouleas}% <-this % stops a space
\affiliation{Department of Electrical Engineering, Duke University, Durham, NC, 27708 USA}% <-this % stops a space

% ABSTRACT
\begin{abstract}
A positron-beam interacting with the plasma electrons drives radial suck-in, in contrast to an electron-beam driven blow-out in the over-dense regime, $n_b>n_0$. In a homogeneous plasma, the electrons are radially sucked-in from all the different radii. The electrons collapsing from different radii do not simultaneously compress on-axis driving weak fields. A hollow-channel allows electrons from its channel-radius to collapse simultaneously exciting coherent fields \cite{hollow-channel}. We analyze the optimal channel radius. Additionally, the low ion density in the hollow allows a larger region with focusing phase which we show is linearly focusing. We have shown the formation of an ion-wake channel behind a blow-out electron bubble-wake. Here we explore positron acceleration in the over-dense regime comparing an optimal hollow-plasma channel to the ion-wake channel \cite{nonlinear-ion-wake}. The condition for optimal hollow-channel radius is also compared. We also address the effects of a non-ideal ion-wake channel on positron-beam excited fields.
\end{abstract}

\maketitle

%%%%%%%%%%%%%%%%%%%%%%%%%%%%%%%%%%%%%%%%%%%%%%%%%%%%%%%%%%%%%%%%%%%%%
% SECTION - Introduction
%%%%%%%%%%%%%%%%%%%%%%%%%%%%%%%%%%%%%%%%%%%%%%%%%%%%%%%%%%%%%%%%%%%%%
% INTRODUCTION
\section{Introduction}
Acceleration of positron beams is a major challenge for developing a collider using plasma-based acceleration techniques. Hollow-channel plasma is required for a collider-level plasma-based positron acceleration \cite{hollow-channel}. However, no analysis has been done to study the scaling laws of the hollow-channel properties based upon the positron beam properties. Similarly, the preparation of hollow-channels with properties suited for positron acceleration is still an ongoing research. In this paper we present a preliminary analysis of the scaling laws and application of the ion-wake channel for positron-beam driven wakefield acceleration. The ion-wake channel is a cylindrical soliton left behind in the plasma by a bubble wake train \cite{nonlinear-ion-wake}. 

Positron beam interacts with a homogenous plasma (density, $n_0$) unlike an ultrashort electron beam (peak density, $n_{be}$, radius, $r_{be}$). The plasma electrons see the repulsive force of the electron beam space-charge and are blown-out radially, $\frac{\operatorname d^2}{\operatorname d\xi^2}r = \frac{1}{r} \frac{1}{2\pi\beta_b^2} \frac{n_{be}(\xi)}{n_0} \pi \left(\frac{r_{be}(\xi)}{c/\omega_{pe}}\right)^2$ (at a fixed $\xi = c\beta_bt - z$, where $\omega_{pe}=\sqrt{\frac{4\pi n_0e^2}{m_e}}$ and $\frac{c}{\omega_{pe}}$ is the plasma skin-depth). We look for radial plasma-electron dynamics at time $t$, just behind the current location of the drive-beam propagating at $c\beta_b$. This is done using an appropriate coordinate $\xi$ such that the current longitudinal beam location is $c\beta_b t$ and $c\beta_b t - z$ is longitudinally behind the beam. The electrons coherently collapse back to the axis under the influence of the plasma ion restoring force; executing cylindrical plasma oscillations and driving a large on-axis longitudinal wakefield. The regions where the electrons are in the blow-out phase, unshielded plasma ions are left behind; quite conveniently exciting the focussing-phase of the transverse fields for an electron beam. 

%%%%%%%%%%%%%%%%%%%%%%%%%%%%
% FIGURE - Homogenous plasma 
%%%%%%%%%%%%%%%%%%%%%%%%%%%%
% Homogenous plasma - positron-beam excited wakefields
\begin{figure}[!htb]
   \centering
   \includegraphics*[width=3.35in]{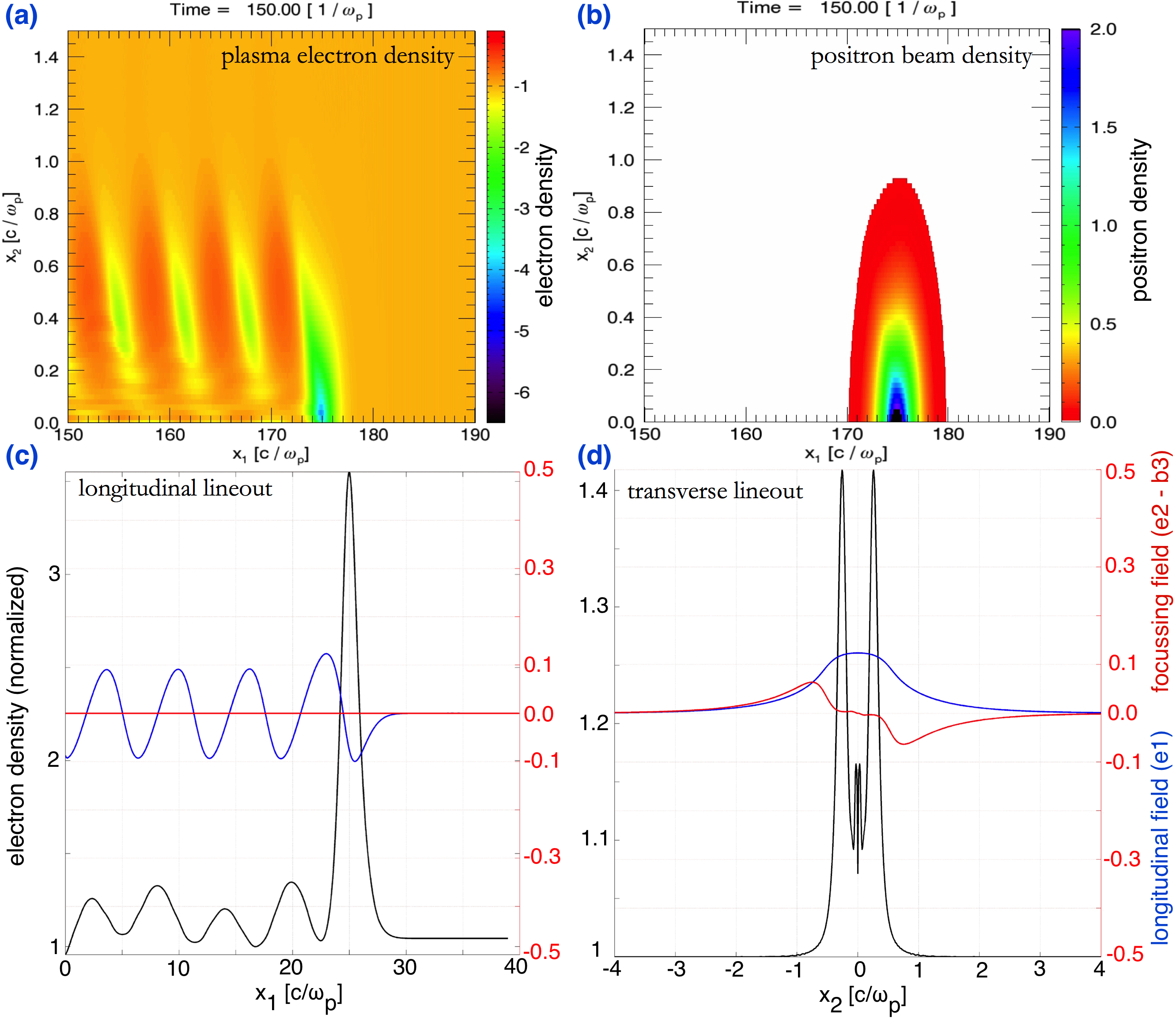}
   \caption{ {\bf Positron-beam driven wakefields in a homogeneous plasma} (a) electron density in 2D cylindrical space at $t=150\omega_{pe}^{-1}$ (b) positron-beam in 2D cylindrical space (c) on-axis longitudinal ($x_1$) line-out: electron density (black), longitudinal field (blue), focussing field (red) (d) transverse ($x_2$) line-out at the peak longitudinal field. }
   \label{Fig1:Homogeneous-plasma-positron-wakefields}
\end{figure}

The space-charge of an ultrashort positron beam ($n_{bp}$, $r_{bp}$), on the contrary, sucks in the plasma electrons and they collapse to the axis, $\frac{\operatorname d^2}{\operatorname d\xi^2}r = - \frac{1}{r} \frac{1}{2\pi\beta_b^2} \frac{n_{bp}(\xi)}{n_0} \pi \left(\frac{r_{bp}(\xi)}{c/\omega_{pe}}\right)^2$. However, there are 2 major problems with the suck-in wakefields in a homogeneous plasma. (i) The plasma electrons are sucked-in from different radii; with the radially closer ones experiencing a larger force and radially farther ones a smaller force (it is interesting to contrast this with an electron beam interacting with the plasma electrons). So, the collapse to the axis is not coherent and different rings of electrons arrive at different times. Due to the lack of optimal on-axis electron compression the wakefields are weak (note the on-axis compression in Fig.\ref{Fig1:Homogeneous-plasma-positron-wakefields}(a)). (ii) When the sucked-in electrons execute radial oscillations, the regions where the electrons are not in the on-axis compression phase has excess unshielded ions. The regions with ions are de-focussing for positrons (positrons to be accelerated are aligned to be on-axis or at a small displacement from the wake axis). The phase where both acceleration and focussing occurs is limited (note the weak focusing force at the location of peak accelerating field in Fig.\ref{Fig1:Homogeneous-plasma-positron-wakefields}(d)). 

%%%%%%%%%%%%%%%%%%%%%%%%%%%%%%%%%%
% FIGURE - Hollow-channel plasma 
%%%%%%%%%%%%%%%%%%%%%%%%%%%%%%%%%%
% Hollow-channel plasma - positron-beam excited wakefields
\begin{figure}[!htb]
   \centering
   \includegraphics*[width=3.35in]{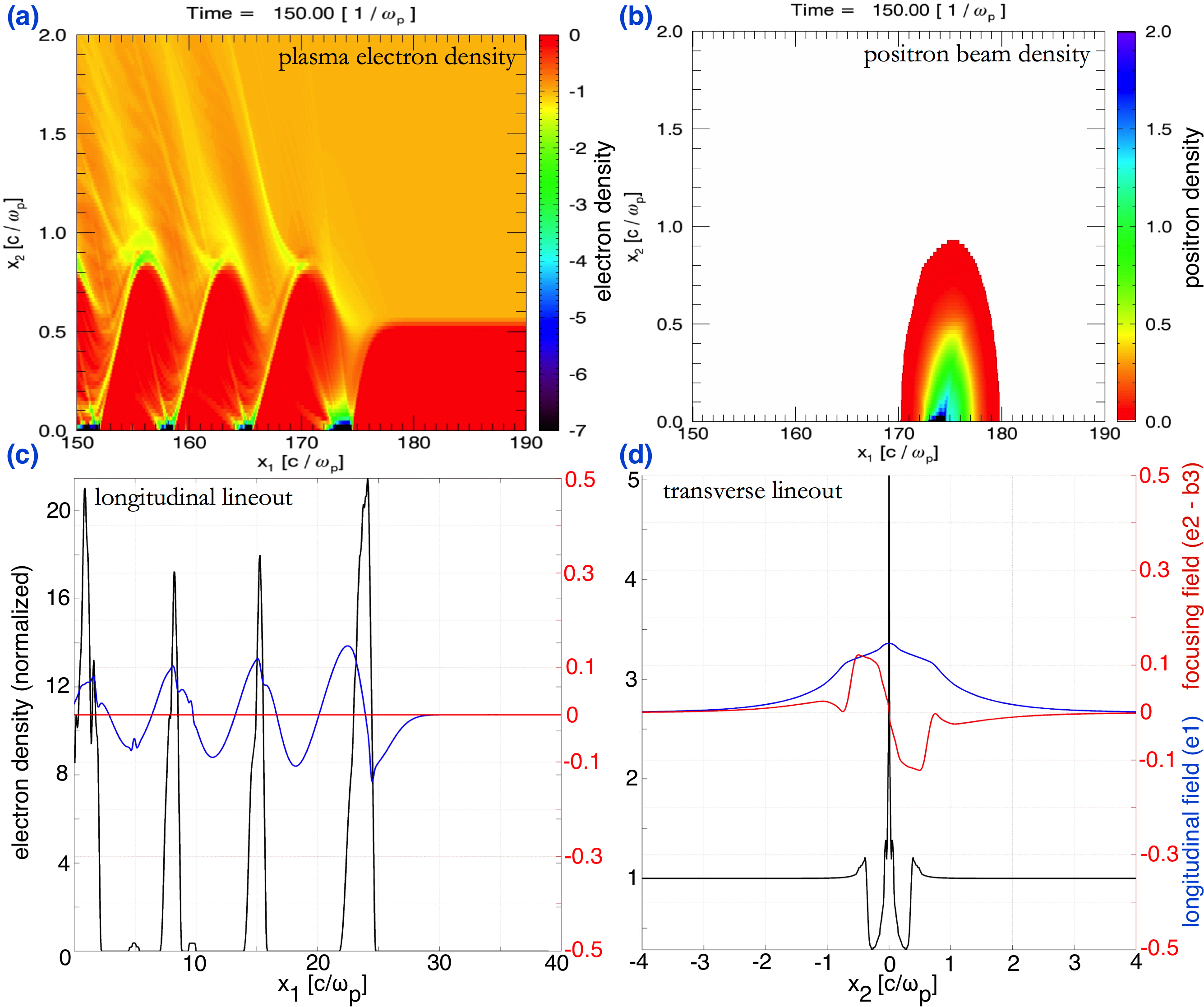}
   \caption{ {\bf Positron-beam driven wakefields in a hollow-plasma channel with $r_{ch}=0.55\frac{c}{\omega_{pe}}$}, with panels equivalent to Fig.\ref{Fig1:Homogeneous-plasma-positron-wakefields} }
   \label{Fig2:Hollow-plasma-positron-wakefields}
\end{figure}

%%%%%%%%%%%%%%%%%%%%%%%%%%%%%%%%%%%%%%%%%%%%%%%%%%%%%%%%%%%%%%%%%%%%%%%%%%%%%%%%%%%%%%
% SUB-SECTION - PIC simulations - comparison of Hollow-channel and Homogeneous plasma
%%%%%%%%%%%%%%%%%%%%%%%%%%%%%%%%%%%%%%%%%%%%%%%%%%%%%%%%%%%%%%%%%%%%%%%%%%%%%%%%%%%%%%
% Compare Homogeneous vs Hollow-channel - PIC simulations
\subsection{Positron-wakefields - Homogeneous vs hollow-channel plasma - Simulations}
Using representative example 2$\frac{1}{2}$-D OSIRIS PIC simulations \cite{osiris-collaboration} we show the differences between the characteristics of a positron-beam excited wakefields. We use Gaussian-shaped positron-beam in all dimensions with $\sigma_z=1.5\frac{c}{\omega_{pe}}$, $\sigma_r=0.3\frac{c}{\omega_{pe}}$ and $\gamma_b=38,000$ in a homogenous plasma, $n_0$ Fig.\ref{Fig1:Homogeneous-plasma-positron-wakefields} and a hollow-channel of radius $r_{ch}$ Fig.\ref{Fig2:Hollow-plasma-positron-wakefields} with radial density profile $n_0(r)=\mathcal{H}(r-r_{ch})$ (where, $\mathcal{H}$ is the Heaviside step function) in cylindrical coordinates to compare the wakefields. We setup a moving window ($40 \times 16 \frac{c}{\omega_{pe}}$) at the speed-of-light which is also almost the speed of the beam ($\beta_b=1$). The plasma consists of pre-ionized fixed-ions. We resolve the smallest spatial scale, $c/\omega_{pe}$ with 20 cells in longitudinal dimension and 50 cells in the transverse dimension (simulation-space of $800\times 800$-cells). The time-step is chosen according to the Courant condition to minimize the numerical dispersion. 

It can be clearly seen in the 2-D density compared in Fig.\ref{Fig1:Homogeneous-plasma-positron-wakefields}(a) and \ref{Fig2:Hollow-plasma-positron-wakefields}(a) that whereas hollow-channel plasma develops regions of on-axis electron compression the homogeneous density plasma has phase-mixing with no distinct compression regions. By comparing the positron beam profile in Fig.\ref{Fig1:Homogeneous-plasma-positron-wakefields}(b) and \ref{Fig2:Hollow-plasma-positron-wakefields}(b) it can be seen that whereas the positron-beam in a hollow-channel sees strong effects of the transverse field, in homogeneous plasma the beam is close to the initialized shape. The transverse wakefields in the region of peak longitudinal fields are compared in Fig.\ref{Fig1:Homogeneous-plasma-positron-wakefields}(d) and \ref{Fig2:Hollow-plasma-positron-wakefields}(d) and in a homogenous plasma the fields are almost negligible in the region close to the axis where the beam with $\sigma_r=0.3\frac{c}{\omega_{pe}}$ is located. So, in this case the transverse wakefields developed by the positron beam in a homogeneous plasma are not usable for transporting the beam. On comparing the longitudinal wakefields we observe that even though the amplitude of the accelerating fields are almost identical in Fig.\ref{Fig1:Homogeneous-plasma-positron-wakefields}(c) and \ref{Fig2:Hollow-plasma-positron-wakefields}(c), the on-axis electron density compression in a hollow-channel plasma is more than 5 times higher than in the homogeneous plasma. The shape of the longitudinal field is much more steepened in \ref{Fig2:Hollow-plasma-positron-wakefields}(c) showing higher non-linearity or greater electron bunching.

We have further examined the effect of pinching of positron-beam on the fields excited in a hollow-channel. We do this by comparing the positron beam in this case with $\gamma_b=38,000$ with a beam of $\gamma_b=10^6$ (much more rigid, $[B\rho]=\gamma\beta\frac{m_ec}{e}$). The results are not shown here but even though for $\gamma_b=10^6$-case the beam-shape is retained the fields excited in the two cases are equal in magnitude and spatial profile.
 
%%%%%%%%%%%%%%%%%%%%%%%%%%%%%%%%%%%%%%%%%%%%%%%%%%%%%%%%%%%%%%%%%%%%%
% SECTION - Optimal condition for hollow-channel plasma
%%%%%%%%%%%%%%%%%%%%%%%%%%%%%%%%%%%%%%%%%%%%%%%%%%%%%%%%%%%%%%%%%%%%%
% Optimal conditions
\section{Hollow-channel - Optimal conditions and Scaling-laws}

To determine the optimality of the hollow-channel properties for developing optimal wakefields (highest peak accelerating field with linear off-axis focusing fields) for positron acceleration, we have to determine the time duration within which the plasma electrons from the channel-edge collapse to the axis. This collapse time then dictates the radius of the channel assuming that the plasma electrons collapse to the axis almost at the speed-of-light. 

The equation of motion of the plasma electron rings at radius, $r$ from the axis, $\frac{\operatorname d^2}{\operatorname d\xi^2}r \propto - \frac{1}{r} n_{bp}(\xi) r_{bp}^2(\xi)$ (where $\frac{d}{d \xi} = '$) is a non-linear second-order differential equation of the form, $r'' = f(r,r',\xi)$ where $f$ is not linear in $r$. We make an important yet simplifying assumption about the positron-beam properties, $n_{bp}(\xi)$ and $r_{bp}(\xi)$: they do not change significantly during the entire interaction of the positron-beam with the hollow-channel over its full length. So, we can drop the dependence on $\xi$ and assume that these properties are constant, $n_{bp}(\xi) = n_{bp}\rvert_{t=0}$ and $r_{bp}(\xi) = r_{bp}\rvert_{t=0}$. With this assumption the non-linear second-order differential equation simplifies to its {\it special case} which has analytical solutions, $r'' = f(r,r')$. We can solve this by setting $v(\xi)=r'(\xi)=\frac{\operatorname dr}{\operatorname d\xi}$, so $v'(\xi)=- \mathcal{C} \frac{1}{r(xi)}$ where $\mathcal{C}=\frac{1}{2\pi\beta_b^2} \frac{n_{bp}}{n_0} \pi \left(\frac{r_{bp}}{c/\omega_{pe}}\right)^2$. If we assume that $v[r(\xi)]$ is an invertible function such that we can obtain, $\hat{v}(r) = v[\xi(r)]$. Then, $\frac{\operatorname d\hat{v}}{\operatorname dr} = \frac{\operatorname dv}{\operatorname d\xi}\frac{\operatorname d\xi}{\operatorname dr}=\frac{v'}{r'}=\frac{v'}{v}$. Substituting for $v'$, we obtain, $\frac{\operatorname d\hat{v}}{\operatorname dr} = -\mathcal{C}\frac{1}{r v(\xi)}$ and $v(\xi)\operatorname d\hat{v} = v\operatorname dv = -\mathcal{C}\frac{\operatorname dr}{r}$. Upon integrating, $v^2 = -2\mathcal{C} ~ \mathrm{ln}(r) + \mathcal{I}_1$. From the initial condition, $v = \frac{\operatorname dr}{\operatorname d\xi}\rvert_{t=0} \simeq 0 \rightarrow r\rvert_{t=0} = r_{ch}$. This gives a differential equation, $\frac{\operatorname dr}{\operatorname d\xi} = \sqrt{2\mathcal{C} ~ \mathrm{ln}(\frac{ r_{ch} }{r})}$ simplifying, $\frac{\operatorname dr}{ \sqrt{\mathrm{ln}(\frac{ r_{ch} }{r})} } = \sqrt{2\mathcal{C}} \operatorname d\xi$. Upon integrating, $- r_{ch} \sqrt{\pi} ~ \mathrm{erf} [\sqrt{\mathrm{ln}( r_{ch} /r)}] = \sqrt{2\mathcal{C}} ~ \xi + \mathcal{I}_2$. Using the initial conditions $\xi=0$ and $t=z=0$ electrons are located at $r = r_{ch}$, the constant of integration is, $\mathcal{I}_2 = -\sqrt{\pi} ~ \mathrm{erf} [\sqrt{\mathrm{ln}( r_{ch} /r_{ch})}] = 0$. So, the expression is $-r_{ch}\sqrt{\pi} ~ ( \mathrm{erf} [\sqrt{\mathrm{ln}( r_{ch} /r)}] ) = \sqrt{2\mathcal{C}} ~ \xi$. When the sucked-in plasma electrons collapse towards the axis, $1/r \rightarrow \infty$ and $\mathrm{erf} [\infty] \rightarrow 1$. So, when the electrons have collapsed to the axis over $\xi=\xi_{coll}$, the time-duration of collapse is $\xi\rvert_{t=0} - \xi_{coll} =  -r_{ch}\sqrt{\frac{\pi}{2}} \frac{1}{\sqrt{\mathcal{C}}}$. We note that there is an anomaly that exists in our problem formulation and the solution because we have not taken into account another force in our equations. This is the space-charge force of the compressing electrons as they collapse to the axis and this force balances the suck-in force of the positron beam. A more careful study of this effect is beyond the scope of this paper. This $\xi_{coll} =  r_{ch}\sqrt{\frac{\pi}{2 \mathcal{C}}} = c\beta_b ~ r_{ch}\frac{ \sqrt{\pi} }{\omega_{pe} \sqrt{ \frac{n_{bp}}{n_0} } r_{pb} } $. Therefore, $t_{coll}=\sqrt{\pi} \frac{ r_{ch}}{\omega_{pe} \sqrt{ \frac{n_{bp}}{n_0} } r_{pb} }$. 

Note that $t_{coll}$ is the time dictated by the beam-properties for the plasma electrons from the channel-radius to collapse to the axis. It has to be matched to the longitudinal compression time. This matching is required to achieve the optimal compression so that ideal longitudinal and transverse fields are excited.

\begin{figure}[!htb]
   \centering
   \includegraphics*[width=3.35in]{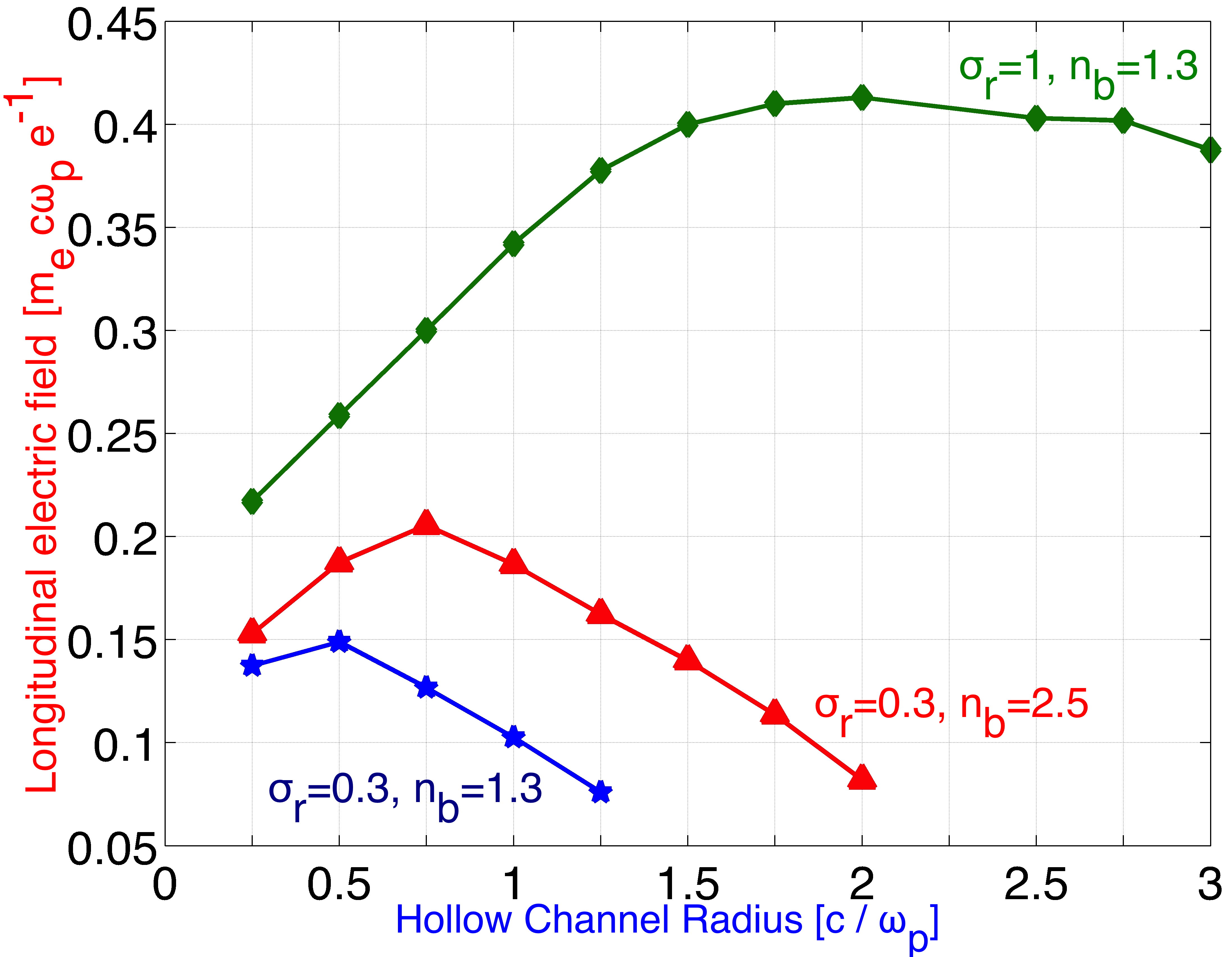}
   \caption{ {\bf Positron-beam driven Longitudinal wakefields vs hollow-plasma channel radius for different beam  properties.}}
   \label{Fig3:Optimal-Hollow-channel-radius}
\end{figure}

To verify the variation of the collapse time and the need to match the channel radius to the time over which compression occurs we use PIC simulations to scan over a few beam parameters and over a set of hollow-channel radius. The results are shown in Fig.\ref{Fig3:Optimal-Hollow-channel-radius} for positron beams with different density and waist parameters as shown in the labels. This is not a comprehensive set of beam parameters so we cannot compare it directly to the solution above. However comparing the 3 curves we see that change in the beam radius has a more significant effect than the beam-density as expected from the solution above.

%%%%%%%%%%%%%%%%%%%%%%%%%%%%%%%%%%
% FIGURE - Radius-vs-Amplitude 
%%%%%%%%%%%%%%%%%%%%%%%%%%%%%%%%%%

\begin{figure}[!htb]
   \centering
   \includegraphics*[width=3.35in]{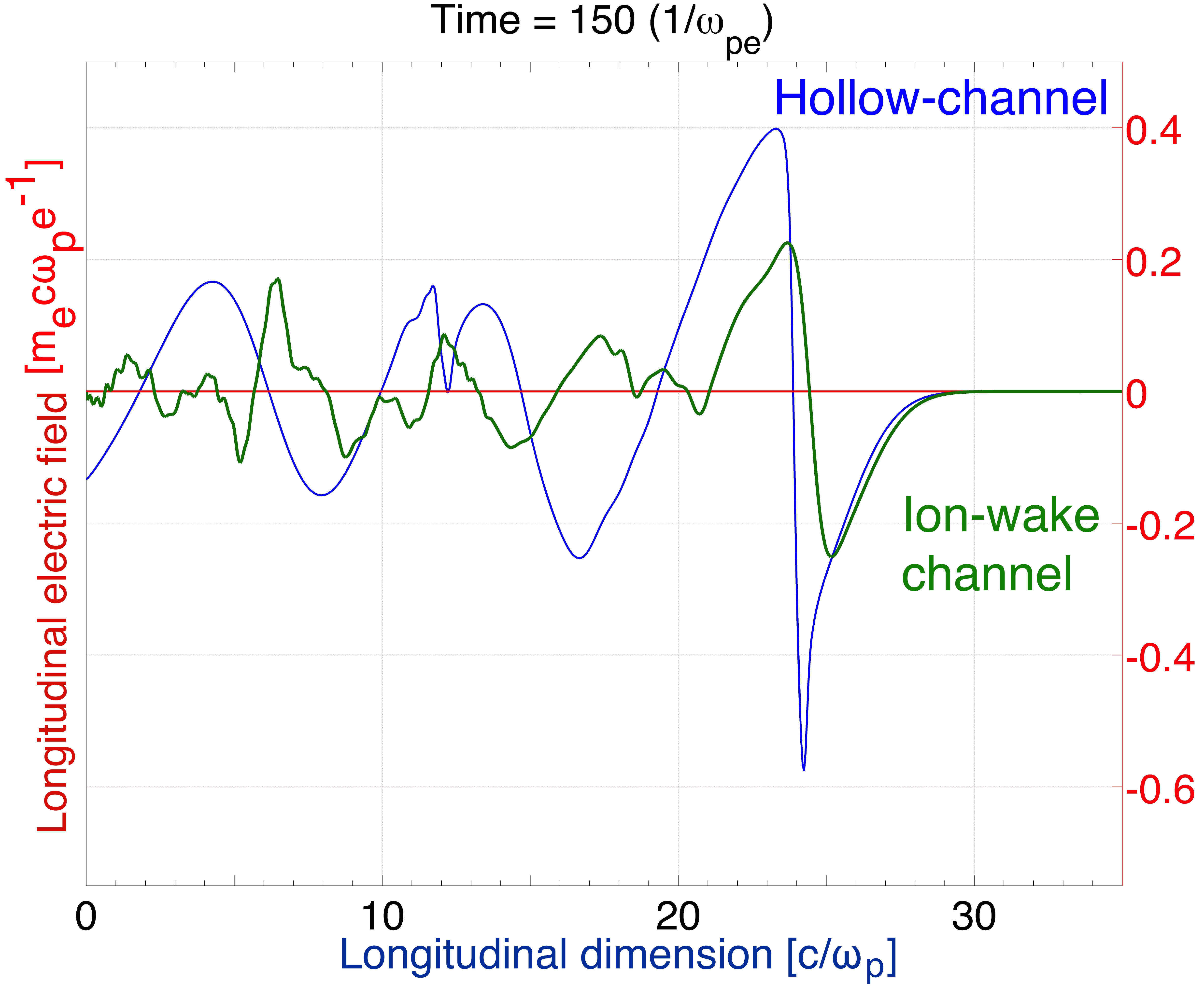}
   \caption{ {\bf Positron-beam driven Longitudinal wakefields - hollow-channel vs ion-wake channel.} Hollow-channel (blue) vs ion-wake channel (green) of channel radius, $2.5\frac{c}{\omega_{pe}}$. }
   \label{Fig4:Ion-wake-hollow-comparison}
\end{figure}

We next compare the properties of the positron-beam (same beam parameters as in Fig.\ref{Fig1:Homogeneous-plasma-positron-wakefields}) driven wakefields excited in an ideal hollow-channel to an ion-wake channel which is excited behind an electron-wakefield bubble-wake train \cite{nonlinear-ion-wake}. In Fig.\ref{Fig4:Ion-wake-hollow-comparison} we show the longitudinal profile of the positron-beam driven accelerating field in an ideal channel (blue) vs in an ion-wake channel (green), both of radius $2.5\frac{c}{\omega_{pe}}$. The ion-wake channel is initialized with an on-axis ($5n_0$, width=$0.3\frac{c}{\omega_{pe}}$) and channel-edge ($4n_0$, width=$1.0\frac{c}{\omega_{pe}}$) density spikes. It is seen that the amplitude of the accelerating field is reduced in an ion-wake channel by about half in this case but the spatial profile is quite similar. Thus it might be quite interesting to explore positron acceleration in the ion-wake channel that is created by the bubble-shaped electron wakefield train. 

%%%%%%%%%%%%%%%%%%%%%%%%%%%%%%%%%%
% FIGURE - Ion-wake vs Hollow
%%%%%%%%%%%%%%%%%%%%%%%%%%%%%%%%%%

\section{Discussion}

We have shown that the positron beam-driven wakefields in a hollow-channel plasma are useful for transporting and accelerating positron beams in a controlled and reliable manner in the non-linear density excitation compression regime. Any collider-level acceleration technique has to be able to support beams of various densities and radii corresponding to different stages of the accelerator. We have explored the scaling of the hollow-channel properties with the positron-beam parameters. We have also shown that ion-wake channels formed behind an electron-wakefield bubble-wake train may be utilized instead of ideal hollow-channels for plasma-based positron acceleration. 

\section{Acknowledgment}
We acknowledge the OSIRIS collaboration for the particle-in-cell (PIC) code \cite{osiris-collaboration}. Work supported by the US Department of Energy under DE-SC0010012 and the National Science Foundation under NSF-PHY-0936278. We acknowledge the {\it Chanakya} server at Duke university.

\newpage
%%%%%%%%%%%%%%%%%%%%%%%%%%%%%%%%%%
% REFERENCES
%%%%%%%%%%%%%%%%%%%%%%%%%%%%%%%%%%

\end{document}